\documentclass[sigconf]{acmart}
\usepackage{enumitem}
\setlist[itemize]{leftmargin=*, itemsep=0pt, topsep=0pt}
\usepackage{graphicx}
\graphicspath{ {./graph/} }
\usepackage{svg}
\usepackage{multirow}
\usepackage{amsmath}
\usepackage{subfigure}
\usepackage{booktabs} 
\usepackage{tabularx} 
\usepackage{xcolor} 
\usepackage{makecell} 
\usepackage{hyperref}
\usepackage{microtype}
\usepackage{balance}

\newcommand{\MSE}{MSE }
\newcommand{\ATL}{ATL }

\AtBeginDocument{%
  }

\copyrightyear{2025}  
\acmYear{2025}  
\setcopyright{acmlicensed}
\acmConference[CIKM '25]{Proceedings of the 34th ACM International Conference on Information and Knowledge Management}{November 10--14, 2025}{Seoul, Republic of Korea.} 
\acmBooktitle{Proceedings of the 34th ACM International Conference on Information and Knowledge Management (CIKM '25), November 10--14, 2025, Seoul, Republic of Korea} 
\acmDOI{10.1145/3746252.3761561} 
\acmISBN{979-8-4007-2040-6/2025/11}

\begin{document}

\title{Climber: Toward Efficient Scaling Laws for Large Recommendation Models}


\author{Songpei Xu}
\email{xusongpei@corp.netease.com}
\affiliation{%
  \institution{NetEase Cloud Music}
  \city{Hangzhou}
  \country{China}
}

\author{Shijia Wang}
\authornote{Corresponding author.}
\email{wangshijia1@corp.netease.com}
\affiliation{%
  \institution{NetEase Cloud Music}
  \city{Hangzhou}
  \country{China}
}

\author{Da Guo}
\email{guoda@corp.netease.com}
\affiliation{%
  \institution{NetEase Cloud Music}
  \city{Hangzhou}
  \country{China}
}

\author{Xianwen Guo}
\email{guoxianwen@corp.netease.com}
\affiliation{%
  \institution{NetEase Cloud Music}
  \city{Hangzhou}
  \country{China}
}

\author{Qiang Xiao}
\authornote{Corresponding author.}
\email{hzxiaoqiang@corp.netease.com}
\affiliation{%
  \institution{NetEase Cloud Music}
  \city{Hangzhou}
  \country{China}
}

\author{Bin Huang}
\email{huangbin02@corp.netease.com}
\affiliation{%
  \institution{NetEase Cloud Music}
  \city{Hangzhou}
  \country{China}
}

\author{Guanlin Wu}
\email{wuguanlin03@corp.netease.com}
\affiliation{%
  \institution{NetEase Cloud Music}
  \city{Hangzhou}
  \country{China}
}

\author{Chuanjiang Luo}
\email{luochuanjiang03@corp.netease.com}
\affiliation{%
  \institution{NetEase Cloud Music}
  \city{Hangzhou}
  \country{China}
}

\renewcommand{\shortauthors}{Songpei Xu et al.}

\begin{abstract}
Transformer-based generative models have achieved remarkable success across domains with various scaling law manifestations. However, our extensive experiments reveal persistent challenges when applying Transformer to recommendation systems: (1) Transformer scaling is not ideal with increased computational resources, due to structural incompatibilities with recommendation-specific features such as multi-source data heterogeneity; (2) critical online inference latency constraints (tens of milliseconds) that intensify with longer user behavior sequences and growing computational demands.
We propose Climber, an efficient recommendation framework comprising two synergistic components: the model architecture for efficient scaling and the co-designed acceleration techniques. Our proposed model adopts two core innovations: (1) multi-scale sequence extraction that achieves a time complexity reduction by a constant factor, enabling more efficient scaling with sequence length; (2) dynamic temperature modulation adapting attention distributions to the multi-scenario and multi-behavior patterns. Complemented by acceleration techniques, Climber achieves a 5.15$\times$ throughput gain without performance degradation by adopting a "single user, multiple item" batched processing and memory-efficient Key-Value caching.

Comprehensive offline experiments on multiple datasets validate that Climber exhibits a more ideal scaling curve. To our knowledge, this is the first publicly documented framework where controlled model scaling drives continuous online metric growth (12.19\% overall lift) without prohibitive resource costs. Climber has been successfully deployed on Netease Cloud Music, one of China's largest music streaming platforms, serving tens of millions of users daily. 
\end{abstract}




\maketitle


\section{Introduction}
Scaling laws, initially explored in language models \cite{kaplan2020scaling,hoffmann2022training}, establish predictable relationships between model performance and key factors such as model size and training data volume.
For instance, Kaplan et al. \cite{kaplan2020scaling} demonstrated that Transformer-based language models \cite{vaswani2017attention} follow power-law improvements in perplexity as model parameters and token counts increase. Similar trends have been observed in vision models \cite{cherti2023reproducible,zhai2022scaling} and multimodal models \cite{bai2023qwen,touvron2023llama}, where the dimensions of scaling model and the diversity of data directly correlate with the performance of downstream tasks. 

Generative recommendation has emerged as the most promising new technical paradigm for enabling scaling laws in recommender systems. We posit that its practical implementation unfolds in a phased manner, where the current stage is primarily dedicated to adapting the Transformer architecture to recommender systems — with the core objective of establishing powerful scaling laws.
Recent research \cite{ardalani2022understanding,guo2024scaling} has validated the effectiveness of scaling laws in recommendation systems, providing valuable insights into model design and resource allocation.
The HSTU model\cite{zhai2024actions} employs hierarchical self-attention mechanisms to model long-term user behavior sequences, achieving better performance than traditional Transformers. Similarly, the MARM model \cite{lv2024marm} introduces memory augmentation to reduce computational complexity, enabling multi-layer sequence modeling with minimal inference costs.
However, these approaches have not adequately addressed the inherent incompatibility between Transformer architectures and recommendation-specific features. Although scaling models through resource expansion remains feasible, this strategy proves inefficient for real-world industrial deployment.
Furthermore, the interplay between key scaling factors---sequence length, model depth, and heterogeneous user behaviors---remains underexplored in traditional recommender systems\cite{yang2024cascading,wang2024sparsity,wang2025enhanced}, leading to suboptimal resource allocation and decreased returns on scaling efforts.



\begin{figure}[t]
  \centering
  \includegraphics[width=\linewidth]{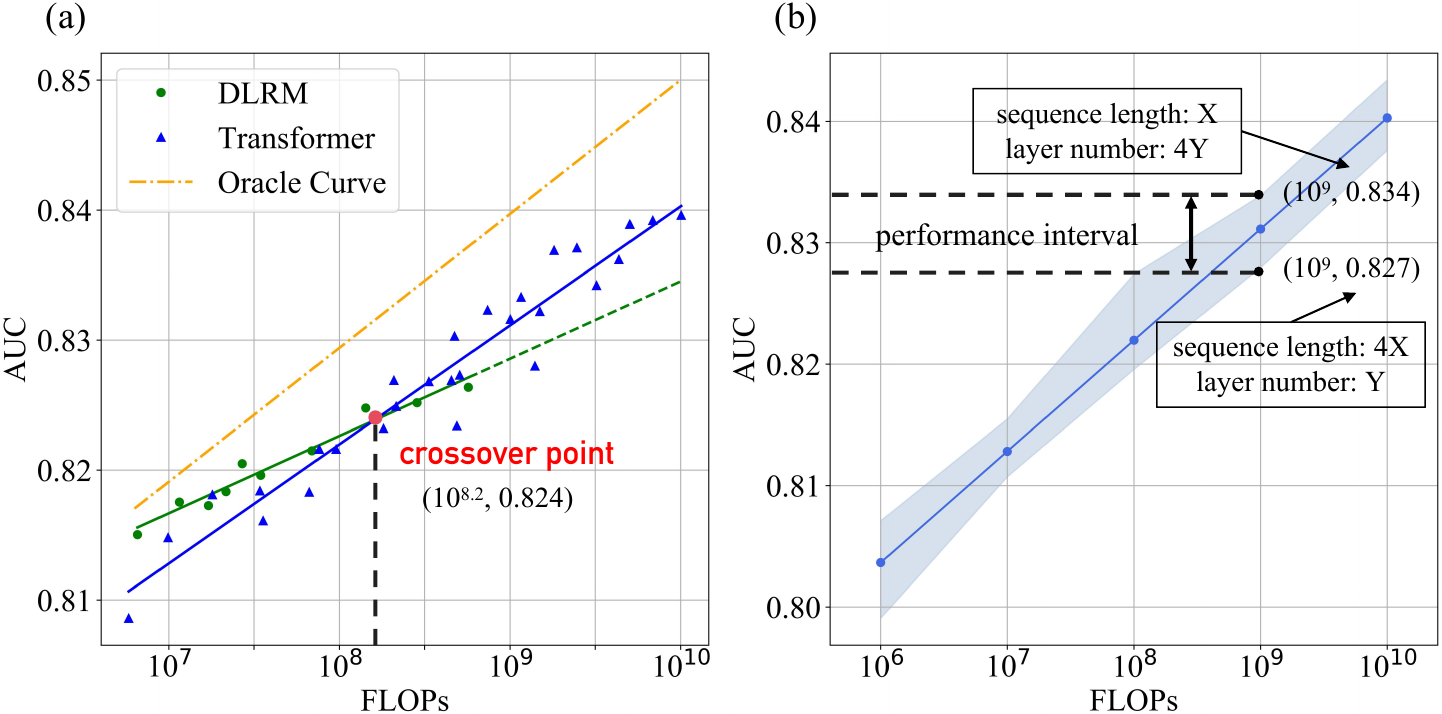}
  \caption{(a) Scalability: DLRM vs. Transformer. The left part of crossover point indicates that Transformer underperforms DLRM when FLOPs are limited. The right part of crossover point highlights the increasing computational demands as Transformer performance improves. The green dashed line indicates DLRM's predicted AUC performance extending beyond observed data points through extrapolation.
  (b) 
  Toy Example: Simulated AUC curve with performance interval across varying sequence length and layer number.
  }
  \label{fig:intro}
  \Description{}
  \vspace{-4mm}
\end{figure}

\renewcommand{\thefootnote}{\fnsymbol{footnote}} 
Inspired by the DeepSeek series \cite{liu2024deepseek1,liu2024deepseek2,bi2024deepseek3}, which has significantly enhanced the efficiency of large language model (LLM) development and reduced computational resource costs, we aim to address the following question: 
\textbf{\textit{how can we efficiently scale up the recommendation model at a substantially lower cost?}}
To gain insights, we conducted industrial-scale analysis on two mainstream models, the Deep Learning Recommendation Model (DLRM)\cite{mudigere2022software} and the Transformer model.
In Figure \ref{fig:intro}(a), we present the scaling curves for both DLRM\footnotemark and Transformer, and an oracle curve is included to represent an ideal scaling curve, characterized by a higher starting point and a larger slope. In Figure \ref{fig:intro}(b), we present an AUC curves derived from simulations of various combinations of sequence lengths and layer number, and introduce the concept of "performance interval", which represents the range of AUC variation for the model under the equivalent FLOPs.
\footnotetext{The baseline model that has been implemented for online deployment.}
However, our findings reveal that some issues still remain when applying Transformer in recommendation systems:
\begin{itemize}
\item \textbf{Transformer's degraded performance under FLOPs constraints}: As shown in Figure \ref{fig:intro}(a), the FLOPs corresponding to the crossover point are $10^{8.2}$. Using this FLOPs value as the boundary, the performance comparison between DLRM and Transformer shows different trends. Transformer model outperforms traditional architectures such as DLRM when FLOPs exceed $10^{8.2}$ magnitude. However, when FLOPs are less than $10^{8.2}$, Transformer model shows worse performance compared with DLRM. This highlights the pursuit for more efficient models with the oracle curve shown in Figure \ref{fig:intro}(a).
The oracle curve represents a more efficient scaling curve with a larger intercept and slope relative to the other scaling curves,
which enables the model to achieve better performance even when the FLOPs are limited.
\item \textbf{Incompatibility between Transformer and recommendation-specific features}: Unlike NLP's continuous syntactic sequences, recommendation systems process fragmented user behaviors spanning multiple scenarios \cite{zhang2024scaling,hou2022towards,li2023text,hou2023learning,luo2022dual}, causing disordered attention distributions as Transformers struggle to prioritize relevant behaviors in sparse multi-source patterns. Additionally, multi-scenario recommendations face distributional discrepancies where users exhibit divergent behaviors, yet existing approaches treat scenarios as auxiliary features rather than explicit distribution controllers. This incompatibility renders Transformers inefficient compared to specialized architectures like DLRMs, particularly under computational constraints.
\item \textbf{The impact of factor combinations on model performance under equivalent FLOPs}: In recommender systems, factors such as sequence length and layer number significantly influence FLOPs, and different combinations of these factors lead to varying model performance. For instance, there is a performance interval(nearly 1\%) under different combinations at $10^9$ FLOPs, as shown in Figure \ref{fig:intro}(b). Current researches lack comprehensive analysis of how factor combinations impact recommendation model performance, hindering efficient model scaling. 
\end{itemize}
These challenges demonstrate Transformers' inefficiency in recommendation systems, which intensifies when scaling features and model capacity to process longer sequences, resulting in quadratic computational demands and tighter latency constraints \cite{chang2023twin,si2024twin}.

Driven by the above insights, we introduce Climber, a novel framework that rethinks the scaling paradigm for recommendation systems. At its core, Climber integrates two complementary innovations: a recommendation-specific Transformer-based model architecture and the co-designed acceleration techniques. Our proposed model redefines how recommendation systems handle user behaviors by introducing multi-scale sequence extraction, which decomposes user behavior sequences into smaller, fine-grained subsequences. This approach not only reduces computational complexity, but also enables more precise modeling of user interests across diverse scenarios. In addition, this model incorporates dynamic temperature modulation, a mechanism that adaptively adjusts attention scores to account for the varying importance of different behaviors and scenarios. 
On the engineering side, we introduce unified acceleration techniques, which transform traditional "single user, single item" sample organization into a format aligned with actual online requests—"single user, multiple items".
Based on these techniques, forward propagation during training and inference with encoder-level KV cache achieves significant efficiency improvements.
Finally, we investigate the scalability of Climber and the impact of factor combinations on AUC under equivalent FLOPs, providing novel insights about rational resource allocation and the key factors for rapid model scaling.

Our contributions are mainly categorized as follows:
\begin{itemize}
\item 
We present the industrial-scale study of scaling laws in recommendation systems, explicitly quantifying the impact of factor combinations under equivalent FLOPs. This analysis reveals that balanced scaling—alternating sequence and depth expansions—yields both offline and online metric gains.
\item We propose a novel Transformer variant, Climber, which resolves the scaling dilemmas in recommendation systems through multi-scale extraction and adaptive temperature modulation. To our knowledge, the proposed method pioneers sustainable scaling—delivering +12.19\% online metric growth, which is the highest annual improvement in our production system.
\item Unified acceleration techniques boost training and inference by "single user, multiple items" batched processing and block-parallel KV cache. Deployed on Netease Cloud Music, 
these techniques sustains 5.15$\times$ training acceleration, and make our model up to 14.38$\times$ faster than DLRM in the online inference, 
enabling 100$\times$ model scaling without increasing computational resources.
\end{itemize}

\section{RELATED WORK}
Wukong \cite{zhang2024wukong} explored parameter scaling in retrieval models but relied on strong assumptions about feature engineering. HSTU \cite{zhai2024actions} reformulated recommendations as sequential transduction tasks, achieving trillion-parameter models with linear computational scaling via hierarchical attention and stochastic sequence sampling. However, HSTU’s focus on generative modeling left gaps in bridging traditional feature-based DLRMs. Concurrently, MARM \cite{lv2024marm} proposed caching intermediate attention results to reduce inference complexity from ${O(n^2d)}$ to $O(nd)$, validating cache size as a new scaling dimension. While effective, MARM’s caching strategy assumes static user patterns, overlooking real-time behavior shifts.

Techniques to mitigate computational costs have been widely adopted. In NLP, KV caching \cite{liu2024scissorhands,dong2024get} avoids redundant attention computations during autoregressive inference. MARM adapted this idea to recommendations by storing historical attention outputs, enabling multi-layer target-attention with minimal FLOPs overhead. Similarly, HSTU introduced Stochastic Length to sparsify long sequences algorithmically, reducing training costs by 80\% without quality degradation. For advertisement retrieval, Wang et al. \cite{wang2024scaling} designed $R/R^*$, an eCPM-aware offline metric, to estimate online revenue scaling laws at low experimental costs. These works collectively highlight the importance of tailoring efficiency strategies to recommendation-specific constraints, such as high-cardinality features and millisecond-level latency requirements.

\section{Method}
To achieve efficient scaling, we propose a Transformer variant specifically designed for recommender systems. This variant supports scaling along three key dimensions: multi-scale sequence processing, multi-scenario adaptation, and multi-interest modeling.
Furthermore, we demonstrate comprehensive deployment details for our proposed model.

\begin{figure}[t]
  \centering
  \includegraphics[width=\linewidth]{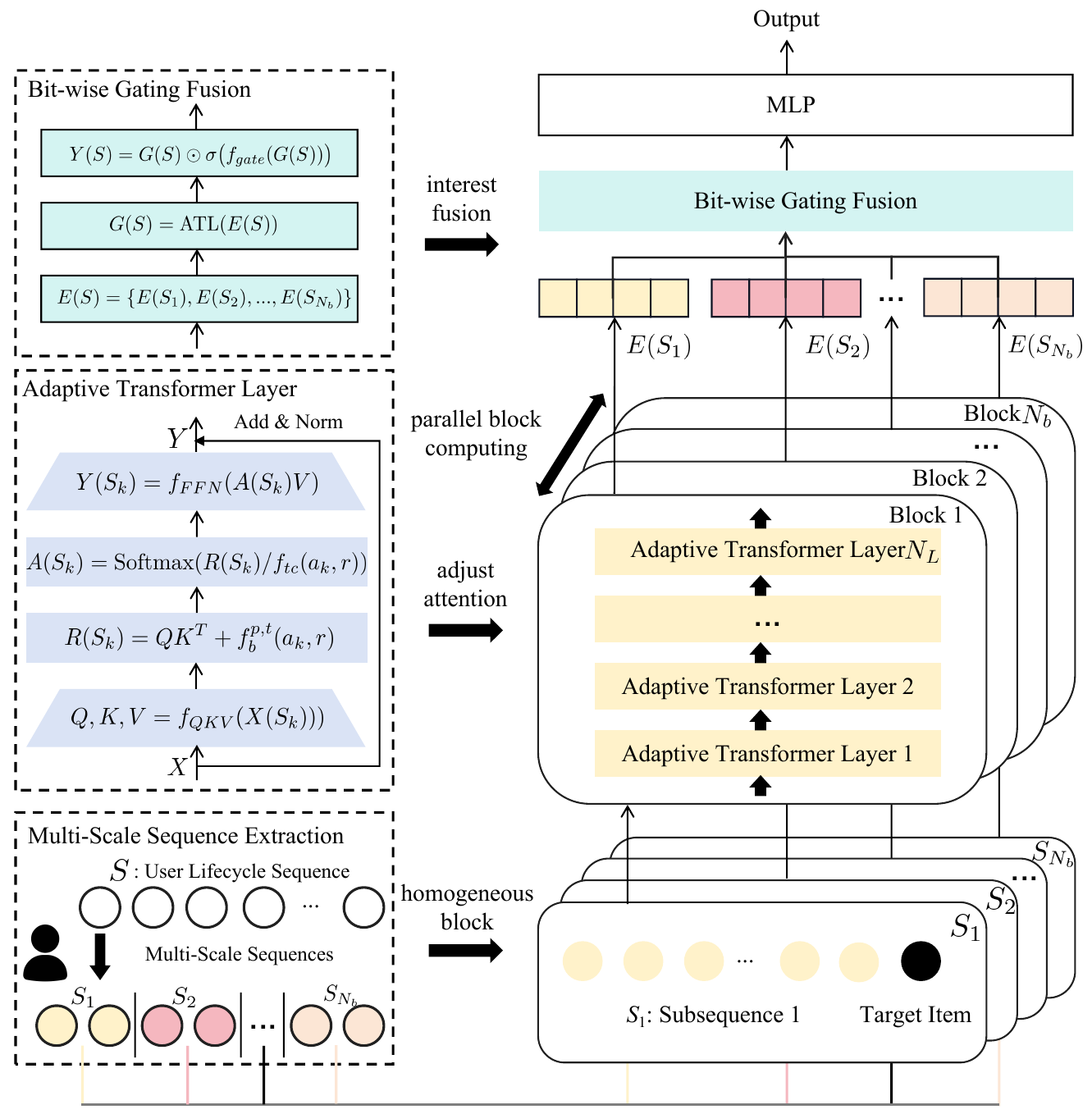}
  \caption{
    Climber Model Architecture.
  }
  \Description{}
  \label{fig:model_arch}
  \vspace{-4mm}
\end{figure}

\subsection{Model Architecture}
\subsubsection{Overall}  

To address the computational complexity and scaling challenges in recommendation systems, we propose the model from a recommendation perspective. 
This model integrates recommendation characteristics into Transformer architecture with resource-aware scalability.
It achieves scaling up from three perspectives: multi-scale sequence, multi-scenario, and multi-interest.
Our model comprises three modules: Multi-scale Sequence Extraction (MSE), Adaptive Transformer Layer (ATL), and Bit-wise Gating Fusion (BGF). Specifically, MSE generates multi-scale sequences from user lifecycle sequence. These multi-scale sequences represent different types of subsequences. 
Each subsequence is processed by a corresponding block composed of stacked ATLs for interest extraction.
Additionally, we extend the time span of important subsequences to cover the user's entire lifecycle.
\ATL adopts an adaptive temperature coefficient to adjust the attention distribution in multi-scenario.
Finally, BGF aggregates the representations from adaptive Transformer blocks to produce a unified output, enabling multi-interest fusion between multi-scale sequences through bit-wise gating mechanism.
Figure \ref{fig:model_arch} illustrates the detailed workflow.

\subsubsection{\textbf{Multi-Scale Sequence Scale-Up}}
We propose a multi-scale sequence extraction (MSE) method for multi-scale sequence scale-up.
This approach reorganizes user sequences based on different strategies, which can be represented by the following formula:
\begin{gather}
S = \{x_1,x_2,...,x_{n_{s}}\} \label{eqn:S} \\
S_k = \text{MSE}(S, a_k) = \{x^{a_k}_{1},x^{a_k}_{2},...,x^{a_k}_{n_k}\} \label{eqn:Sk}
\end{gather}
where $S$ represents the user lifecycle sequence,  $x_i$ denotes $i$-th item ID from the entire item set $\textbf{X}$. $S_k$ represents the $k$-th subsequence extracted from the user lifecycle sequence $S$ based on the extraction strategy $a_k$, and $x^{a_k}_{j}$ indicates the $j$-th item in the subsequence $S_k$. The $n_s$ and $n_k$ denote the length of $S$ and $S_k$, respectively. We assume that there are a total of $N_b$ extraction strategies and $\sum_{k=1}^{N_b} n_k=n$. In our practical application, $n \ll n_s$. This is because the extraction strategy typically retains only the user's positive behaviors. Consequently, the computational complexity under a single Transformer can be reduced from $O(n_s^2d)$ to $O(n^2d)$. We further improve the training process by employing the corresponding Transformer block for each subsequence $S_k$, thereby achieving a time complexity of $O(n_k^2d)$. We design each extraction strategy to extract subsequences of equal length, such that $n_k=n/N_b$. In the case of fully serial operations, this results in a time complexity of $O(n^2d/N_b)$. Therefore, even when $N_b=2$, we can still achieve substantial training acceleration.
Under full parallelism with adequate computational resources, the complexity reduces to $O(\text{max}(n_k)^2d)$, and the complexity only depends on the longest subsequence length. This guarantees training efficiency preservation when incorporating sequences shorter than $n_k$, thereby enabling progressive multi-scale sequence scale-up. Our extraction strategies include business-driven sequences (e.g., click/like/share), model-filtered sequences, etc.
In summary, \MSE reduces the computational complexity and transform user lifecycle sequence into multi-scale sequences. These enhancements improve the efficiency and scalability of the recommendation system.

\subsubsection{\textbf{Multi-Scenario Scale-Up}}
The Softmax activation function, which normalizes the attention scores, plays a pivotal role in the Transformer architecture. Specifically, the attention mechanism is normalized by $\frac{QK^T}{\sqrt{d_{k}}}$ and subsequently multiplied by $V$. The division by $\sqrt{d_{k}}$ ensures that the distribution of the attention matrix aligns with that of $Q$ and $K$ \cite{hinton2015distilling}. 
However, we generate multi-scale sequences based on different extraction strategies from user lifecycle sequence, and apply corresponding Transformer block to each subsequence.
The distribution of each subsequence also exhibits significant differences across different scenarios.
A single scaling factor $\sqrt{d_{k}}$ is insufficient to accommodate the diverse requirements of all Transformer blocks corresponding to multi-scale sequences in mutli-scenario \cite{he2018determining}.
To further refine the attention distribution within each Transformer block, we introduce an adaptive temperature coefficient for each layer of each block.
We refer to this change as Adaptive Transformer Layer (ATL) , which can be mathematically expressed as follows:
\begin{equation}
\begin{gathered}
Q,K,V =f_{QKV}(X(S_k)), \\
R(S_k) = QK^{T} + f_{b}^{p,t}(a_k, r),  \\ 
A(S_k) = \text{Softmax}(R(S_k)/f_{tc}(a_k, r)), \\
Y(S_k) = f_{FFN}(A(S_k)V).
\end{gathered}
\label{eqn:ATL}
\end{equation}
Compared to conventional Transformer layer, we introduce an adaptive temperature coefficient and adjust relative attention bias from a recommendation perspective. Here, $X(S_k)\in \mathbb{R}^{s \times d}$, $R(S_k)\in \mathbb{R}^{h \times s \times s} $, $A(S_k) \in \mathbb{R}^{h \times s \times s}$ and $Y(S_k) \in \mathbb{R}^{s \times d}$ represent layer input, raw attention matrix, normalized attention matrix, and layer output, respectively. $s$, $h$, $d$ represent sequence length, head number and feature dimension, respectively.
$f_{QKV}(X(S_k))$ is used to derive query, key, and value matrices from input $X(S_k)$.
$f_{b}^{p,t}(a_k, r)$ denotes relative attention bias \cite{zhai2024actions, raffel2020exploring} that incorporates positional ($p$) and temporal ($t$) information.
$f_{tc}(a_k, r)$ denotes a function that derives the temperature coefficient.
It is important to note that both extraction strategy $a_k$ and recommendation scenario $r$ influence the relative attention bias and the temperature coefficient.
$f_{FFN}$ processes the attention-weighted value matrix through a feedforward neural network (FFN) to produce the final output of the layer. This approach is inspired by the multi-scenario and multi-behavior characteristics of recommendation systems. Unlike HSTU's fixed temperature coefficient\cite{zhai2024actions}, our adaptive temperature coefficient allows for more flexible attention weighting, addressing the limitations of the fixed scaling factor $\sqrt{d_{k}}$ in capturing the inherence of diverse behaviors and scenarios.

\subsubsection{\textbf{Multi-Interest Scale-Up}}
However, when user lifecycle sequence is separated into multi-scale sequences, there is a lack of interaction between them \cite{xiao2020deep,li2019multi}. Therefore, we propose a bit-wise gating fusion module that integrates information across the $N_b$ subsequences corresponding to our $N_b$ blocks to achieve multi-interest scale-up. Specifically, each block generates an output vector $E(S_k)$ through each adaptive Transformer block, and these $N_b$ output vectors are concatenated as $E(S) \in \mathbb{R}^{N_b \times d}$. The concatenated vectors are then processed through a new ATL, followed by a sigmoid activation function to achieve bit-level gating. Finally, the vectors pass through a subsequent network to produce the final output score. The bit-wise gating fusion module can be expressed as:
\begin{equation}
\begin{gathered}
    E(S) = \{E(S_1),E(S_2),...,E(S_{N_b})\}, \\
    G(S) = \text{ATL}(E(S)), \\
    Y(S) = G(S) \odot \sigma\big(f_{gate}(G(S))\big).
\end{gathered}
\end{equation}
where $\sigma$ is sigmoid activation and $f_{gate}$ represents a squeeze-and-excitation module \cite{hu2018squeeze}, which ensure the identity between the input and output dimensions to dynamically adjusts the contribution of each element in $G(S) \in \mathbb{R}^{N_b \times d}$.
$f_{gate}(G(S)) \in \mathbb{R}^{N_b \times d}$ and $Y(S) \in \mathbb{R}^{N_b \times d}$ represent the bit-wise attention matrix and the output of fusion module, respectively.
$\text{ATL}$ (Adaptive Transformer Layer in Equation \ref{eqn:ATL}) is used to calculate the similarity between different blocks to interact the different interest from subsequences. In contrast, the ATL in fusion function doesn't incorporate relative attention bias, and the temperature coefficient is solely determined by the recommendation scenario.
The attention operation of ATL on $N_b$ blocks can be regarded as field-wise interactions \cite{hu2018squeeze,huang2019fibinet}, which facilitate information exchange at the feature level. Our method enhances multi-interest fusion by adding bit-wise interactions. This allows the model to capture precise relationships among sequences. As a result, the model's ability to understand multi-scale sequences is significantly improved.
It is worth mentioning that although this attention mechanism has a complexity of $O(N_b^2d)$, the number of extraction strategy $N_b$ is much smaller than the dimension $d$ of the feature vector in the fusion stage. Thus the computational complexity of bit-wise gating fusion module is relatively low.


\subsection{Deployment}\label{deployment}
Our acceleration deployment involves two phases: offline training and online serving.
In the offline training phase, user interaction logs are recorded as "single user, single item" pattern. These raw logs are compressed and archived as "single user, multiple items" pattern.
Here, "single user, single item" denotes a log entry recording one atomic interaction (e.g., click/purchase) between a user and an item; "single user, multiple items" aggregates a user's historical interactions into a single record with multiple items, enabling batch processing of user-item feature computation.
The "single user, multiple items" pattern uses full-visible masks between each candidate item and the entire history, and also employ diagonal masks for inter-item isolation between candidate items.
This compression mechanism significantly reduces sample volume while providing 5.15$\times$ acceleration for model training. 
In the online serving phase, motivated by M-FALCON \cite{zhai2024actions}, the system first generates multi-layered key-value (KV) cache vectors from user features, then fetches candidate item features from the feature server, and finally computes attention-based interactions between the item features and cached KV representations.
Notably, the model adopts a "single user, multiple items" data pattern across both offline training and online serving phases, thus utilizing KV cache to accelerate user feature computation.
In addition, we implement operator fusion to merge sequential operations (e.g., embedding lookup, attention layers) into unified computational kernels to reduce frequent global memory accesses, and integrate FlashAttention's matrix tiling operations to optimize memory utilization\cite{dao2022flashattention}.
These designs enhance computational efficiency through cache utilization while maintaining prediction accuracy, thereby delivering personalized recommendation services with improved user satisfaction.

\section{Experiments}
In this section, we detail the offline and online experiments conducted on real industrial data to evaluate our proposed method, addressing the following four research questions:
\begin{itemize}
\item RQ1: How does Climber perform in offline evaluation compared to state-of-the-art (SOTA) models?
\item RQ2: How does Climber demonstrate superior scalability compared to DLRM and Transformer?
\item RQ3: How can we allocate resources to scale up our model by considering the impact of different factor combinations on AUC under equivalent FLOPs? 
\item RQ4: How does Climber perform in industrial systems?
\end{itemize}


\subsection{Experimental Setting}
\subsubsection{Dataset}
To validate the effectiveness of our method in recommendation systems, we construct a dataset using real user behavior sequence as the primary feature. The dataset is used to predicts different user actions on the candidate items based on historical interactions. Important user behaviors include full-play, like, share and comment. 
Additionally, we evaluated our model on three recommendation datasets, \textbf{Spotify}\cite{brost2019music}, \textbf{30Music}\cite{turrin201530music} and \textbf{Amazon-Book}\cite{mcauley2015image}. 
Table \ref{tab:Dataset} presents the number of users, items, and interactions for the four processed datasets. To protect data privacy, we only present statistical data for our recommendation scenarios, and a special processing has been applied to the Industrial dataset. As a result, the data volume shown in the table is lower than the actual amount. However, it is clear that the scale of our industrial dataset still significantly exceeds that of the other datasets. This substantial data volume provides a robust foundation for conducting scaling experiments.


\begin{table}[t]
\centering
\caption{Dataset Statistics.}
\label{tab:Dataset}
\begin{tabular}{@{}lcccccc@{}}
\toprule
 & Spotify & 30Music & Amazon-Book & Industrial \\
\midrule
\#User & 0.16M & 0.02M & 0.54M & \textbf{>40M} \\
\#Item & 3.7M & 4.5M & 0.37M & \textbf{>6M} \\
\#Interaction & 1.2M & 16M & 1.09M & \textbf{>1B} \\
\bottomrule
\vspace{-6mm}
\end{tabular}
\end{table}

\begin{table*}[htbp]
\centering
\caption{Evaluation of Methods on Public/Industrial Datasets.}
\begin{tabular}{c|cc|cc|cc|cc}
\hline
\multicolumn{1}{c|}{} & \multicolumn{2}{c|}{Spotify} & \multicolumn{2}{c|}{Amazon-Book} & \multicolumn{2}{c|}{30Music} & \multicolumn{2}{c}{Industrial} \\
 & AUC & LogLoss & AUC & LogLoss & AUC & LogLoss & AUC & LogLoss \\ \hline
DLRM(Baseline) & 0.7606 & 0.5761 & 0.7842 & 0.5541 & 0.8927 & 0.2012 & 0.8216 & 0.7067 \\ \hline
DIN & 0.7557 & 0.5803 & 0.7796 & 0.5580 & 0.8861 & 0.2044 & 0.8158 & 0.7109 \\
TWIN & 0.7589 & 0.5772 & 0.7831 & 0.5563 & 0.8903 & 0.2019 & 0.8203 & 0.7078 \\ 
Transformer & 0.7621 & 0.5735 & 0.7836 & 0.5557 & 0.8930 & 0.2007 & 0.8214 & 0.7074 \\ 
HSTU & 0.7626 & 0.5722 & 0.7869 & 0.5520 & 0.8938 & 0.1982 & 0.8217 & 0.7053 \\ \hline
Climber (-ATL, -BGF) & 0.7635 & 0.5710 & 0.7873 & 0.5518 & 0.8944 & 0.1978 & 0.8221 & 0.7045 \\
Climber (-BGF) & 0.7655 & 0.5694 & 0.7881 & 0.5510 & 0.8950 & 0.1972 & 0.8225 & 0.7034 \\ 
\underline{Climber} & \underline{0.7663} & \underline{0.5687} & \underline{0.7887} & \underline{0.5501} & \underline{0.8986} & \underline{0.1957} & \underline{0.8230} & \underline{0.7029}
\\
\textbf{Climber-large} & \textbf{0.7702} & \textbf{0.5666} & \textbf{0.7914} & \textbf{0.5472} & \textbf{0.9035} & \textbf{0.1916} & \textbf{0.8398} & \textbf{0.6911} \\ \hline
\textbf{\%Improve} & \textbf{+1.26\%} & \textbf{-1.64\%} & \textbf{+0.91\%} & \textbf{-1.24\%} & \textbf{+1.20\%} & \textbf{-4.77\%} & \textbf{+2.21\%} & \textbf{-2.20\%} \\ \hline
\end{tabular}
\label{tab:ASTRO}
\end{table*}

\subsubsection{Compared methods}
This section details the experimental settings for each model in the Industrial Dataset.
\begin{itemize}
\item DLRM: 
this model leverages lifelong user behavior sequences and complex feature interactions, which has been deployed in our online systems. In experiments, the sequence length is fixed at 2000.
\item DIN: 
DIN captures user interests via a target attention mechanism modeling interactions between user historical behaviors and the target item. 
In experiments, sequence length is set to 1000.
\item TWIN: 
this model aligns GSU and ESU to enhance consistency in modeling long-term user behavior. 
In experiments, GSU and ESU sequence length are set to 2000 and 1000, respectively.
\item Transformer: 
A Transformer-based model for sequence modeling. In experiments, a one-stage Transformer encoder processes behavior sequences with a fixed length of 2000.
\item HSTU: 
The feature-level sequence is temporally reorganized via chronological ordering of item-action pairs and processed through the HSTU model to predict target item-specific user actions. In experiments, the sequence length is fixed at 2000.
\item Climber Series: 
The previous method confines the sequence length to 2000 within a fixed time window without behavior filtering. With multi-scale sequence extraction, the new approach extends behavior sequences to the entire user lifecycle through business logic-driven strategies, retaining a reduced length of 200. The model maintains 2 layers, consistent with the baseline configuration. In the industrial setting, the Climber-large variant scales to 12 layers and 800 sequence length.
\end{itemize}

\subsection{Overall Performance (RQ1)}
\subsubsection{Performance Comparison}

As shown in Table \ref{tab:ASTRO}, Climber achieves the best performance in four recommendation datasets. Notably, the application scenarios of \textbf{Spotify} and \textbf{30Music} belong to the same domain as our industrial dataset, which is music recommendation systems. In contrast, \textbf{Amazon-Book} differs significantly from our scenario. However, our model still achieves favorable results on this dataset, indicating its potential adaptability to diverse applications.
Next, we focus on comparing the AUC improvements of Climber relative to other methods. 1) as our primary online model, DLRM outperforms DIN and TWIN because DLRM includes a wide range of feature interaction structures beyond lifelong sequence and attention mechanisms. 
2) The Transformer achieves +0.134\% AUC improvement over TWIN. This is because Transformer computes the similarity between all historical items and target item in a single stage.
3) HSTU implements several enhancements to Transformer, achieving an AUC improvement of +0.036\% on our dataset compared to Transformer. However, these enhancements also result in increased computational complexity.
4) Climber reduces computational complexity by sequence extraction and adjusts attention distribution across multi-scenario and multi-behavior through adaptive temperature coefficients. This results in a +0.170\% improvement over DLRM. Furthermore, Climber-large achieves a +2.21\% AUC improvement by scaling up the model, achieving the largest offline gain in the past year.

\subsubsection{Abalation Study}
To evaluate the contributions of each component in our Climber model, we conduct a comprehensive set of experiments across multiple datasets. For illustrative purposes, we focus on the ablation study conducted on the Industrial dataset and select Transformer for comparison with Climber series. By incorporating MSE, Climber (-ATL, -BGF) transforms user lifecycle sequence into multi-scale subsequence blocks. This enhancement results in a positive AUC gain of +0.085\% compared to the Transformer model. Climber (-BGF) further improves the model by introducing an adaptive temperature coefficient. This component adjusts the attention distribution dynamically, leading to an AUC improvement of +0.134\%. Finally, the incorporation of BGF brings about an AUC improvement of +0.195\%. This module integrates user interests represented by different subsequences, emphasizing the importance of interest fusion in recommendation systems. In summary, our model demonstrates strong performance and adaptability based on the offline evaluations across various datasets.

\begin{figure*}[htbp]
  \centering
  \includegraphics[width=\linewidth]{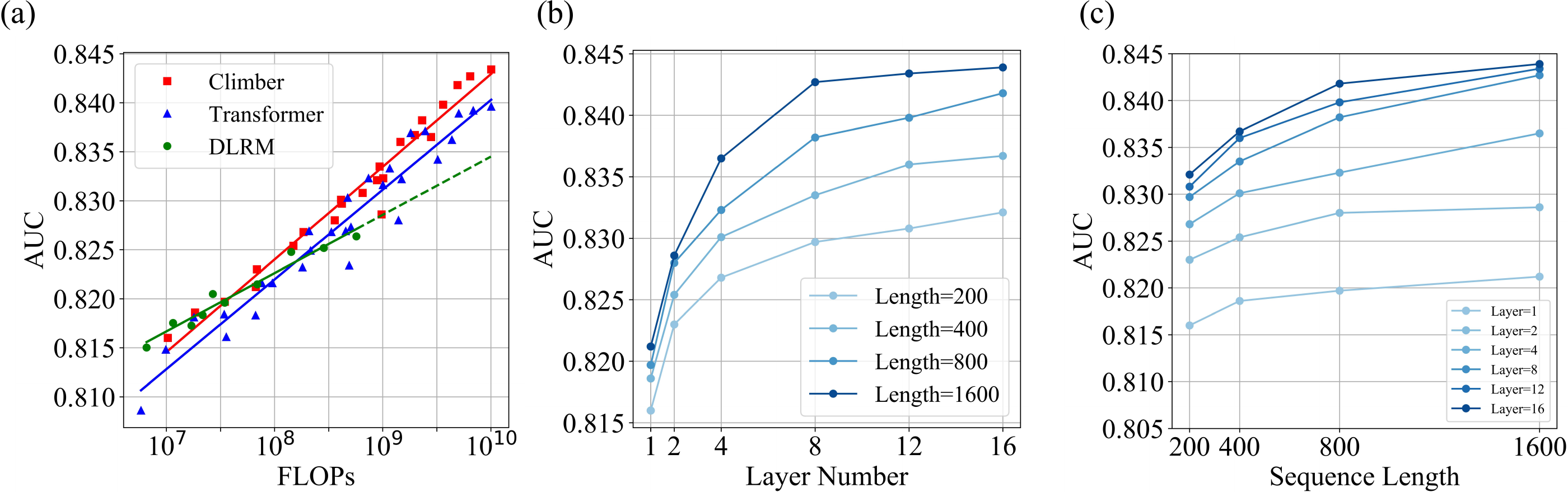}
  \caption{Scaling Curve. (a) Scalability: DLRM vs Transformer vs Climber in Industrial Dataset. (b) Model performance of scaling layer number with constant sequence length. (c) Model performance of scaling sequence length with constant layer number.}
  \Description{}
  \label{fig:scale}
\end{figure*}

\subsection{Scalability (RQ2)}
Before discussing model scalability, we formally define FLOPs as $C \propto s * l$, 
where $s$ represents the sequence length and $l$ represents the layer number in the model. 
In large-scale scenarios, the quadratic computational complexity of attention mechanisms accounts for only a minor proportion of the model's overall FLOPs even with longer sequence\cite{casson2023transformerflops,kaplan2020scaling,hoffmann2022training}.Thus in our computational analysis, we focus solely on the linear part of sequence length $s$ within  $C \propto s * l$ relationship, and the FLOPs can be calculated and validated by specific tools of Tensorflow.
The scaling curves of DLRM, Transformer, and Climber are shown in Figure \ref{fig:scale}(a). Although Transformer can achieve better performance than DLRM when FLOPs exceed $10^9$, its efficiency is notably lower than DLRM between $10^7$ and $10^8$. Compared to the Transformer, Climber exhibits a more ideal scaling curve due to its higher starting point and larger slope. When FLOPs are below $10^{7.5}$, Climber's performance remains weaker than DLRM, but the crossover point shifts to the left, enabling the Climber model to achieve performance transformation more efficiently than Transformer.
In this experiment, the two main factors affecting FLOPs are layer number and sequence length. For the Climber, we illustrate the relationship between model performance and the layer number in Figure \ref{fig:scale}(b) and sequence length in Figure \ref{fig:scale}(c), respectively. When the sequence length is fixed, model performance improves in a manner similar to a power-law with the increase in the layer number; when the layer number is fixed, there is a similar improvement in model performance as the sequence length increases.
Thus, our proposed Climber model exhibits scaling curves in terms of FLOPs, sequence length, and layer number, and has a more efficient scaling curve compared to Transformer.

\subsection{Efficient Allocation (RQ3)}

From Figure \ref{fig:scale}(b, c), it is evident that increasing sequence length and layer number can improve the model's AUC. However, the priority of these two factors lacks discussion. 
Table \ref{tab:performance_comparison} presents the model's AUC for the equivalent FLOPs with different layer number and sequence length. It is clear that under the equivalent FLOPs, the combination of layer number and sequence length can lead to significant changes in the offline testing AUC.
According to $C \propto s*l$, the product of layer number and sequence length remains constant for the equivalent FLOPs. When the FLOPs is $4.11 \times 10^8$, the model achieves the best performance of 0.8301 AUC on a ($400s\times4l$) model; When the FLOPs is $1.01 \times 10^9$, the model achieves the best performance of 0.8335 AUC on a ($400s\times8l$) model.
We observe that expanding a single factor may limit the model's development. Therefore, it is best to consider both layer number and sequence length equally when scaling up the model. For example, with a ($400s\times4l$) model, if we need to increase FLOPs by 4 times, we can choose between ($1600s\times4l$), ($800s\times8l$), and ($400s\times16l$). From Table \ref{tab:performance_comparison}, it can be found that the best choice is ($800s\times8l$), which jointly expands both factors to increase the model's AUC from 0.8301 to 0.8382.
This conclusion also guides how to allocate resources online. In our practical recommendation systems, usually only one factor is chosen for each iteration. Therefore, when scaling up online, we alternate between increasing sequence length and layer number.

\begin{table}[htbp]
\centering
\caption{
Performance Comparison under Equivalent FLOPs.
}
\label{tab:performance_comparison}
\begin{tabular}{@{}lcccccc@{}}
\toprule
FLOPs & Sequence Length & Layer Number & AUC \\
\midrule
\multirow{4}{*}{$4.11 \times 10^{8}$} & 1600 & 1 & 0.8212   \\
& 800 & 2 & 0.8280   \\
& \textbf{400} & \textbf{4} & \textbf{0.8301}  \\
& 200 & 8 & 0.8297   \\
\midrule
\multirow{4}{*}{$1.01 \times 10^{9}$} & 1600 & 2 & 0.8286  \\
 & 800 & 4 & 0.8323   \\
 & \textbf{400} & \textbf{8} & \textbf{0.8335}  \\
 & 200 & 16 & 0.8321 \\
 \midrule
\multirow{3}{*}{$2.55 \times 10^{9}$} & 1600 & 4 & 0.8365  \\
 & \textbf{800} & \textbf{8} & \textbf{0.8382}   \\
 & 400 & 16 & 0.8367  \\
\bottomrule
\end{tabular}
\end{table}

\begin{table}[htbp]
  \caption{Online Metric Improvement Compared with DLRM.}
  \label{tab:online_metric}
  \begin{tabularx}{1\columnwidth}{@{}ccccc@{}}
    \toprule
    Method & FLOPs & \makecell{Sequence \\ Length} & \makecell{Layer \\ Number} & \makecell{Online Metric}\\ 
    \midrule
    DLRM & $3.45\times 10^7 (6\times)$ & - & - & $+0\%$ \\
    \midrule
    \multirow{8}{*}{Climber} & $5.82\times 10^6 (1\times)$ & $100$ & $1$ & $-4.95\%$ \\
     & $1.84\times 10^7 (3\times)$ & $400$ & $1$ & $-1.31\%$ \\
     & $3.46\times 10^7 (6\times)$ & $800$ & $1$ & $-1.22\%$ \\
     & $4.11\times 10^8 (71\times)$ & $400$ & $4$ & $+3.65\%$ \\
     & $1.01\times 10^9 (174\times)$ & $800$ & $4$ & $+4.29\%$ \\
     & $2.79\times 10^9 (479\times)$ & $1600$ & $4$ & $+7.78\%$ \\
     & $2.31\times 10^9 (397\times)$ & $800$ & $8$ & $+10.68\%$ \\
     & $3.61\times 10^9 (620\times)$ & $800$ & $12$ & \textbf{$+12.19\%$}\\
  \bottomrule
  \vspace{-7mm}
  \end{tabularx}
\end{table}

\subsection{Online A/B Test (RQ4)}
Table \ref{tab:online_metric} summarizes the online A/B test results of our proposed Climber framework.
Consistent with the conclusion that sequence length and layer number are equally important, our model demonstrates online scaling curves for metric and FLOPs by only adjusting sequence length and layer number.
Firstly, Climber with $5.82\times10^6$ FLOPs exhibits a negative metric improvement. When the FLOPs value of Climber (6$\times$) model match that of DLRM (6$\times$), there is only a slight negative metric, indicating that Climber model's efficiency is lower with fewer FLOPs compared with DLRM. Furthermore, when the FLOPs value of Climber (479$\times$) model is $2.79\times10^9$, the online metric improvement reaches +7.78\%. Finally, when the FLOPs value of Climber (620$\times$) model is $3.61\times10^9$, an online metric improvement of +12.19\% is achieved. 
In the online inference stage, Climber achieves significantly lower latency—specifically 2.92$\times$-14.38$\times$ faster per request-than DLRM on varying length sequence and layer number. This acceleration is attained by our acceleration techniques with the equivalent inference budget used by the DLRMs (Section \ref{deployment}), thus we can deploy models that are 100$\times$ more complex.
To the best of our knowledge, Climber is the first recommendation model to display both offline and online scaling curves while maintaining resource balance. Moreover, it achieves +12.19\% metric improvement, representing the largest improvement in the past year.

\section{Conclusion}
We propose Climber—an efficient scaling framework with a specific Transformer variant and co-designed acceleration techniques.
Our model effectively reduces computational complexity and breaks scaling dilemma in recommendation systems by addressing multi-scale sequences, multi-scenario, and multi-interest aspects.
This integration enables the model to exhibit superior scalability in offline evaluation compared to both DLRM and Transformer models.
Furthermore, we introduce the acceleration techniques, which employs "single user, multiple items" sample format and an encoder-level KV cache. These techniques enable the deployment of models that are 100$\times$ more complex without increasing prohibitive computing resources. 
The Climber demonstrates a scaling curve online and achieves a 12.19\% improvement in online metric.
In summary, this work adapts the Transformer architecture to recommender systems within constrained resources, establishing a foundation for the next stage of the generative recommendation paradigm. In the future, we will explore more generative techniques in recommender system, which aims to continuously unlock scaling potential.


\bibliographystyle{ACM-Reference-Format}
\balance


\end{document}